\documentclass[useAMS,usenatbib,a4paper]{mn2e}
\usepackage{amsmath}
\usepackage[pdftex]{graphicx}
\usepackage{epstopdf}
\usepackage[english]{babel}
\usepackage{subfigure}
\usepackage{array,multirow}
\usepackage{underscore}
\usepackage{natbib}
\usepackage{float}
\usepackage{amssymb}
\usepackage{placeins}

\newcolumntype{L}[1]{>{\raggedright\let\newline\\\arraybackslash\hspace{0pt}}m{#1}}
\newcolumntype{C}[1]{>{\centering\let\newline\\\arraybackslash\hspace{0pt}}m{#1}}
\newcolumntype{R}[1]{>{\raggedleft\let\newline\\\arraybackslash\hspace{0pt}}m{#1}}

\def\mnras{MNRAS}
\def\apj{ApJ}
\def\aj{AJ}

\def\apjl{ApJL}
\def\apjs{ApJS}

\def\pasp{PASP}
\def\nat{Nature}

\title{The fundamental plane of evolving red nuggets}
\author[L. J. Oldham et al.]{Lindsay Oldham$^{1}$\thanks{E-mail: loldham@ast.cam.ac.uk}, Matthew Auger$^{1}$, Christopher D. Fassnacht$^{2}$, Tommaso Treu$^{3}$,
\and L.V.E. Koopmans$^{4}$, David Lagattuta$^{5}$, John McKean$^{4,6}$, Simona Vegetti$^{7}$\\
$^{1}$ Institute of Astronomy, University of Cambridge, Madingley Road, Cambridge CB3 0HA, UK \\
$^{2}$ Department of Physics, University of California, Davis, 1 Shields Ave. Davis, CA 95616, USA\\
$^{3}$ Department of Physics and Astronomy, UCLA, 430 Portola Plaza, Los Angeles, CA 90095-1547, USA\\
$^{4}$ Kapteyn Astronomical Institute, University of Groningen, P.O. Box 800, 9700 AV Groningen, The Netherlands\\
$^{5}$ Universit\'e de Lyon, Universit\'e de Lyon, CNRS, Centre de Recherche Astrophysique de Lyon UMR5574, F-69230, Saint-Genis-Laval, France\\
$^{6}$ ASTRON, Netherlands Institute for Radio Astronomy, Postbus 2, NL-7990 AA, Dwingeloo, the Netherlands\\
$^{7}$ Max Planck Institute for Astrophyiscs, Karl-Schwarzschild-Strasse 1, D-85740 Garching, Germany\\
}
\pubyear{2016}

\begin{document}
\maketitle
\setcounter{page}{1}

\begin{abstract}
We present an exploration of the mass structure of a sample of 12 strongly lensed massive, compact early-type galaxies at redshifts $z\sim0.6$ to provide further possible evidence for their inside-out growth. We obtain new ESI/Keck spectroscopy and infer the kinematics of both lens and source galaxies, and combine these with existing photometry to construct (a) the fundamental plane (FP) of the source galaxies and (b) physical models for their dark and luminous mass structure. We find their FP to be tilted towards the virial plane relative to the local FP, and attribute this to their unusual compactness, which causes their kinematics to be totally dominated by the stellar mass as opposed to their dark matter; that their FP is nevertheless still inconsistent with the virial plane implies that both the stellar and dark structure of early-type galaxies is non-homologous. We also find the intrinsic scatter of their FP to be comparable to the local value, indicating that variations in the stellar mass structure outweight variations in the dark halo in the central regions of early-type galaxies. Finally, we show that inference on the dark halo structure -- and, in turn, the underlying physics -- is sensitive to assumptions about the stellar initial mass function (IMF), but that physically-motivated assumptions about the IMF imply haloes with sub-NFW inner density slopes, and may present further evidence for the inside-out growth of compact early-type galaxies via minor mergers and accretion. 

\end{abstract}

\section{Introduction}

The discovery that massive, passive galaxies at redshifts $z\sim2$ are much more compact than their present-day counterparts \citep{Daddi2005,Trujillo2006,vanDokkum2008} has led to a picture in which early-type galaxies (ETGs) evolve dramatically in size over the course of their lives. Moreover, the detection of extended outer envelopes surrounding lower-redshift compact ETGs \citep{vanDokkum2010,Oldham2016} implies an important role for dissipationless merging and accretion in evolving these systems towards the present-day size-mass relation. However, though a consensus is now building over the evolution of their luminous structure, their dark halo structure remains elusive. Even in normal (non-compact) ETGs, how the dark halo is affected by baryonic processes such as mergers and accretion is not well understood; simulations suggest that dynamical heating from infalling satellites should displace dark matter from the centre of the halo to larger radii \citep{ElZant2004,Laporte2012}, but this must compete with other processes such as adiabatic contraction during the infall of stellar material and feedback from supernovae and active galactic nuclei \citep{Read2005,Martizzi2013}. Observationally, the picture is also unclear, with the halo structure of ETGs exhibiting a diversity which may depend on environment \citep{OldhamAuger2016b,Newman2015}. Probing the mass structure of low-redshift, partly-evolving massive compact `red nugget' galaxies, where the evolution of the baryonic material is dominated by merging, allows us to isolate this aspect of inside-out growth from other baryonic processes and investigate its impact on the haloes of individual galaxies much more closely.

The mass structure of ETGs \emph{as a population}, on the other hand, has historically been accessed through the fundamental plane (FP) which tightly relates their characteristic size, velocity dispersion and surface brightness \citep{djorgovski,dressler87}. The existence of such a plane follows directly from the assumption that galaxies are virialised -- and, to some extent, homologous -- systems \citep[see e.g.][]{Ciotti1996}, with the small intrinsic scatter indicating a strong degree of regularity in their formation and evolution. However, the fact that the FP is tilted relative to the virial prediction implies some degree of non-homology, with mass-dependent variations in either the luminous matter -- for instance, the stellar initial mass function (IMF), stellar mass structure and stellar dynamics -- or the dark matter -- including the dark halo structure, concentration and the dark matter fraction -- or both; however, it is difficult to disentangle effects of variations in the dark and light structure \citep[see e.g.][]{Trujillo2004, Cappellari2006, LaBarbera2008} so as to extract information on galaxy structure \citep[though the construction of the \emph{mass plane} has helped to discount dynamical non-homology as the main cause of the FP tilt; see][]{Bolton2008}. Constructing the FP for compact galaxies, whose luminosity-weighted velocity dispersions probe the very central regions where the dark matter fraction is expected to be low, provides a way to separate the contributions to the FP tilt from the stars and the halo and so better understand the mass properties of the ETG population. 

In this paper, we present new high signal-to-noise ESI/Keck spectroscopy for the 13 early-type/early-type lenses (EELs) of \citet{Oldham2016}, and combine these data with photometry to probe the mass structure both of individual galaxies and of the population as a whole. The background galaxies of the EELs are massive and compact ETGs, which the combination of lensing magnification and high-resolution imaging data can resolve on $\sim 100$ pc scales, making them the ideal sample with which to probe ETG structural evolution. In Section 2, we introduce the data, the data reduction and the kinematic modelling. Section 3 presents the FP; in Section 4 we construct physical models of the EELs sources in order to set constraints on their dark matter content, and in Section 5 we discuss our findings and conclude. Throughout the paper, we use circularised radii and assume a flat $\Lambda$CDM cosmology with $\Omega_m = 0.3$ and $h = 0.7$.


\section{Data and kinematic modelling}

We observed the 13 EELs using the Echelette Spectrograph and Imager \citep[ESI;][]{ESI} on Keck on the nights of 2013 May 14, 2015 Jan 23 and 2016 July 08, obtaining 1-hour exposures for each system (except J1218+5648, J1605+3811 and J2228-0018, which were observed for 30 minutes, 10 minutes and 10 minutes respectively), using a slit width of 0.75$''$. The data were reduced with a custom-made, python-based pipeline and the wavelength scale calibrated using arc lamp exposures taken on the night. For each system, we extracted spectra over two separate apertures -- one centred on the lens, with width $0.5 ''$, and a second centred on the brightest part of the Einstein ring, with width $0.3 ''$ -- to obtain spectra which maximised the relative signal from the lens and the source respectively. For one system (J1606+2235), the slit did not cover the surface brightness peaks in the source galaxy which meant that we were not able to extract a spectrum of the central regions of the source in this case; we therefore exclude it from our sample. The source galaxy spectra for the 12 remaining systems are presented in Figure 1.

To determine the stellar velocity dispersions, we model each spectrum as the sum of a lens, source and additive continuum component. For the lens and source, we use stellar templates for A, F, G and K stars from the Indo-US Stellar Library of Coud\'{e} Feed Stellar Spectra \citep{valdes}, which we redshift and convolve with a dispersion $\sigma_{model}^2 = \sigma_{true}^2 + \sigma_{inst}^2 - \sigma_{tmp}^2$ where $\sigma_{true}$ is the physical velocity dispersion of the system, $\sigma_{inst} = 20.30 $ kms$^{-1}$ is the instrument resolution and $\sigma_{tmp}$ is the intrinsic resolution of the templates (which is 1.2 $\mathrm{\AA}$ for the Indo-US templates). The continuum is an order-6 polynomial which accounts for the difference in shape between the templates and the true spectrum, and regions where atmospheric absorption dominates the spectrum are masked. We therefore have four free non-linear parameters -- the redshift and velocity dispersion for each of the lens and the source -- and 24 linear parameters -- the weights of each of the nine stellar templates for the source and the lens (4 K stars, 3 G stars and one A and F star each), and the coefficients of the order-6 polynomial -- which we explore using the Markov Chain Monte Carlo (MCMC) package \texttt{emcee} \citep{ForemanMackey2013}. Our kinematic models are also shown in Figure 1 and the kinematics for the sources, together with their photometric properties, are summarised in Table 1. We defer the presentation of the lens galaxy kinematics to a future work. 

We test the robustness of our kinematic inference by repeating the modelling process using the lower-resolution galaxy templates of \citet{Bruzual2003} and find that the typical uncertainty in the velocity dispersion of the source is of order $5 \%$; as this is significantly larger than our statistical uncertainties, we impose this as the uncertainty on all our velocity dispersion measurements. We also check the robustness of our method by modelling the kinematics of (a) the ESI spectra, extracted over apertures of width $1.5''$, centred on the lens (to emulate the SDSS apertures, which are cirular with radius 1.5$''$) and (b) the actual SDSS spectra, and compare our inference on the lens kinematics in each case; the uncertainty indicated by these tests is typically smaller than the uncertainty due to the choice of templates. 

We combine these kinematics with the photometry presented in \citet{Oldham2016} in order to construct both the FP and physical mass models. In that paper, parametric light and lensing mass distributions were used to model the HST/ACS V/I and Keck/NIRC2 $K'$ imaging of the thirteen EEL systems. We refer the reader to that paper for full details on the modelling and results, but summarise the effective radii $R_e$ and effective surface brightnesses $\log I_e$ (defined as the average surface brightness within the effective radius, evaluated in the rest-frame Johnson $V$ band using the photometric models of \citealp{Oldham2016}) in Table 1. 

\begin{figure*}
    \centering
\includegraphics[trim=20 20 20 20,clip,width=0.49\textwidth]{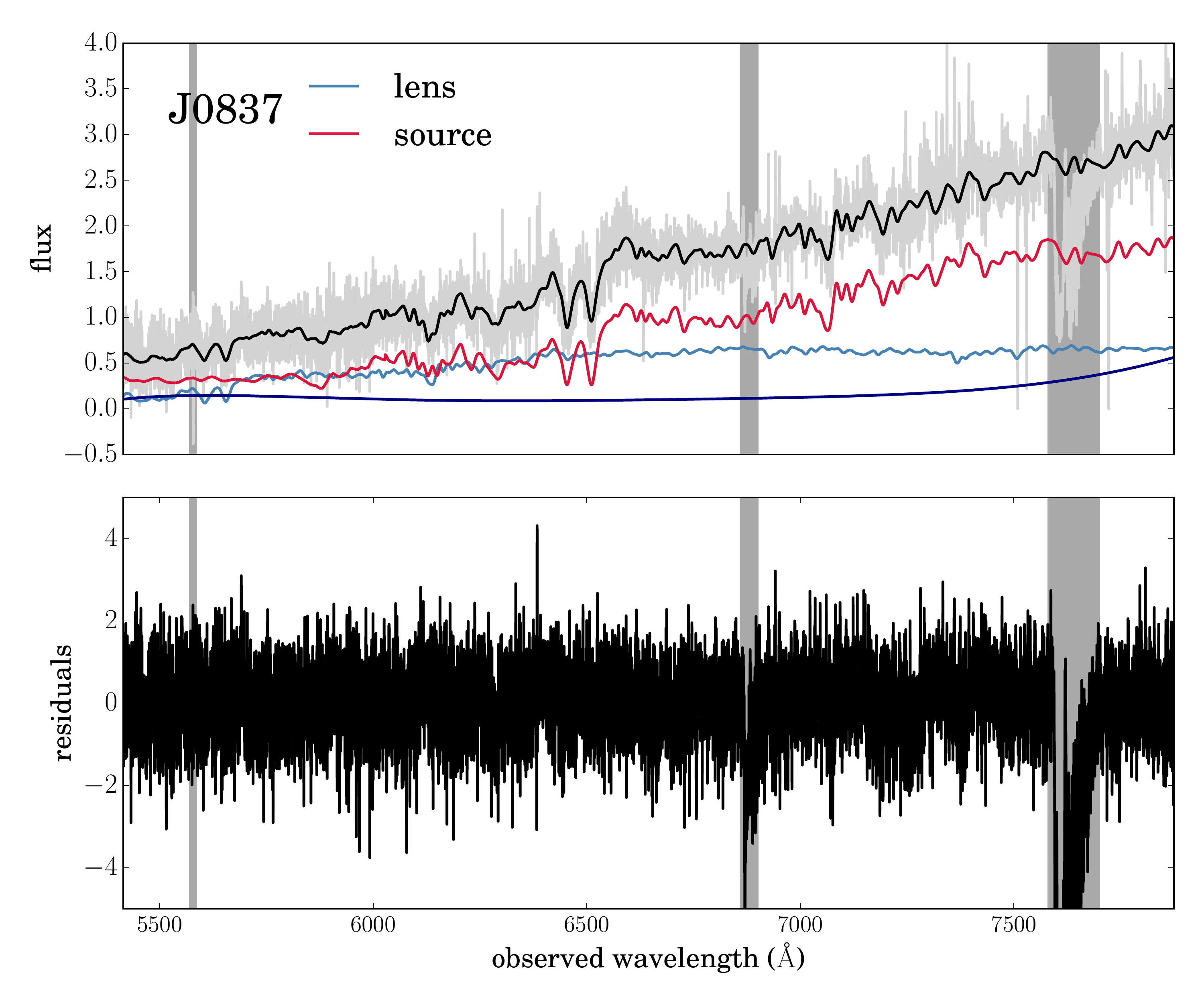}
\includegraphics[trim=20 20 20 20,clip,width=0.49\textwidth]{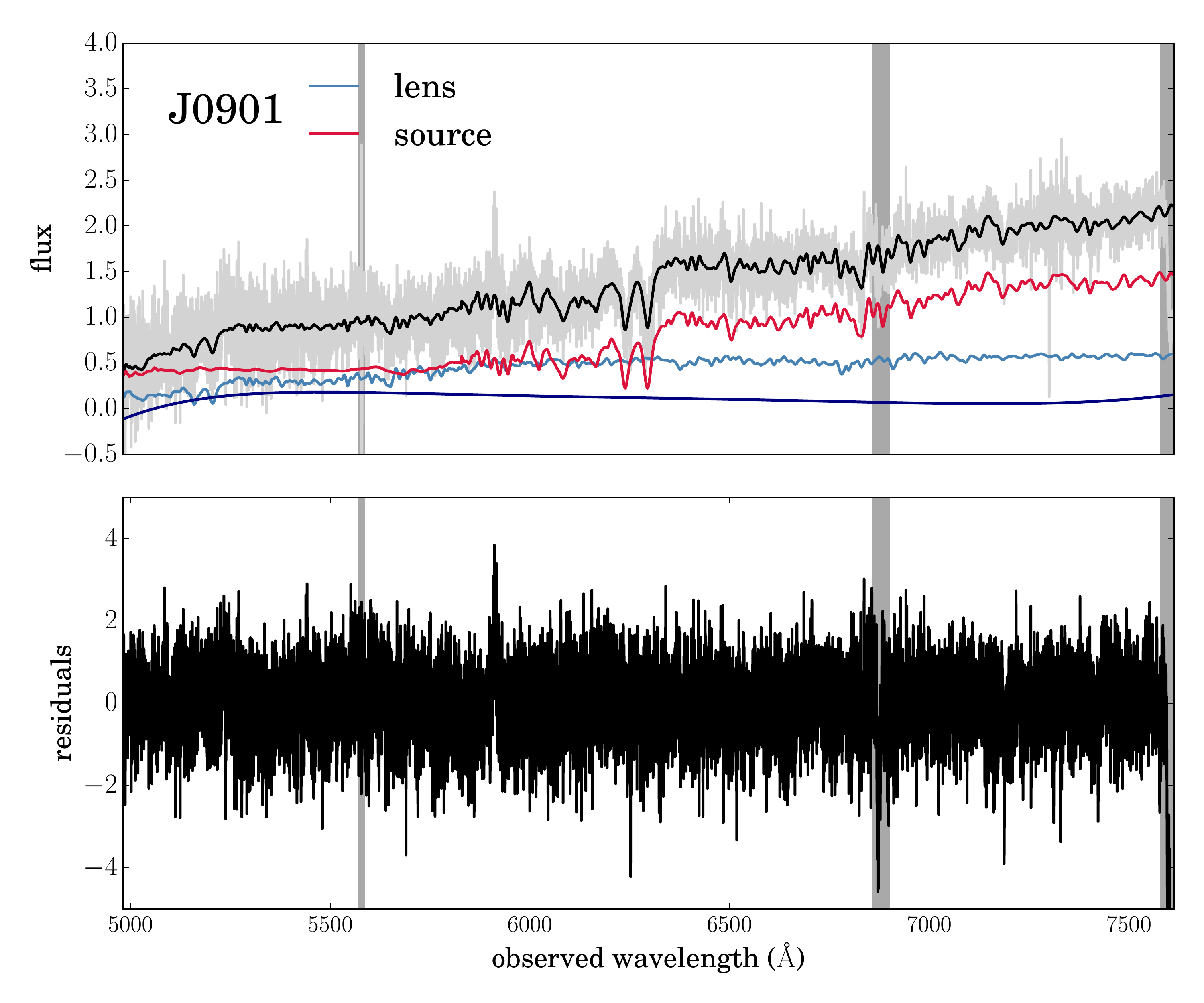}

\includegraphics[trim=20 20 20 20,clip,width=0.49\textwidth]{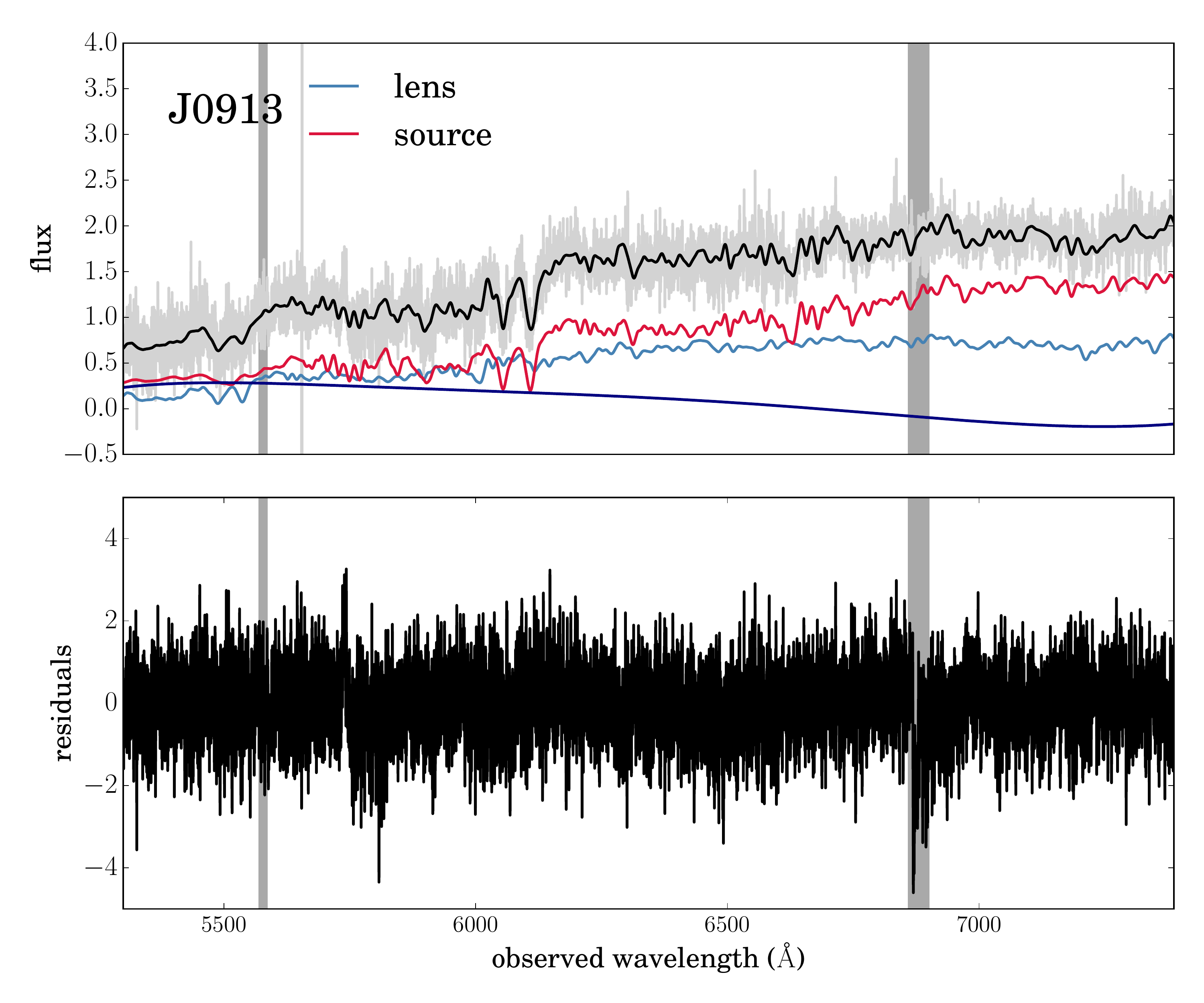}
\includegraphics[trim=20 20 20 20,clip,width=0.49\textwidth]{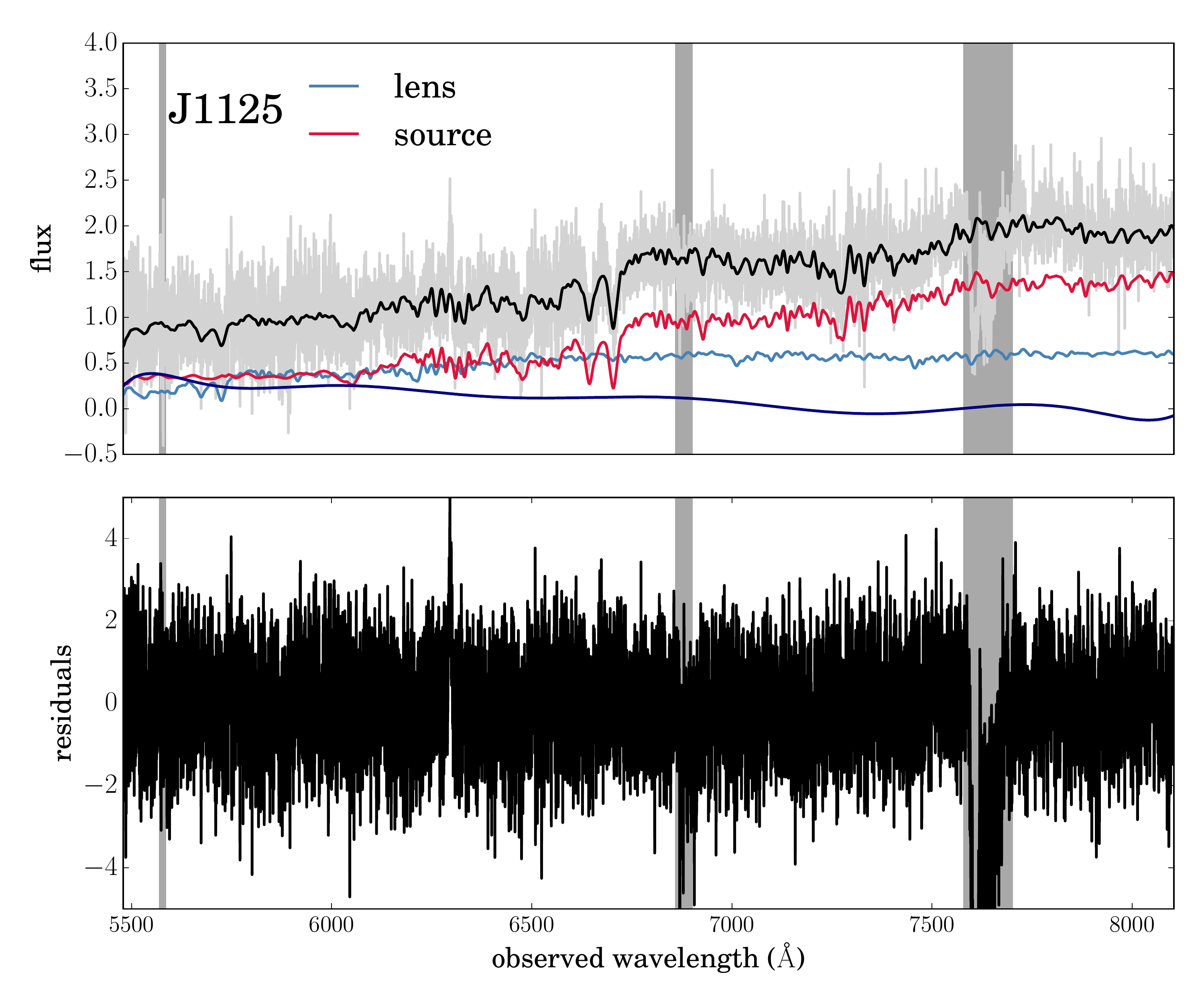}

\includegraphics[trim=20 20 20 20,clip,width=0.49\textwidth]{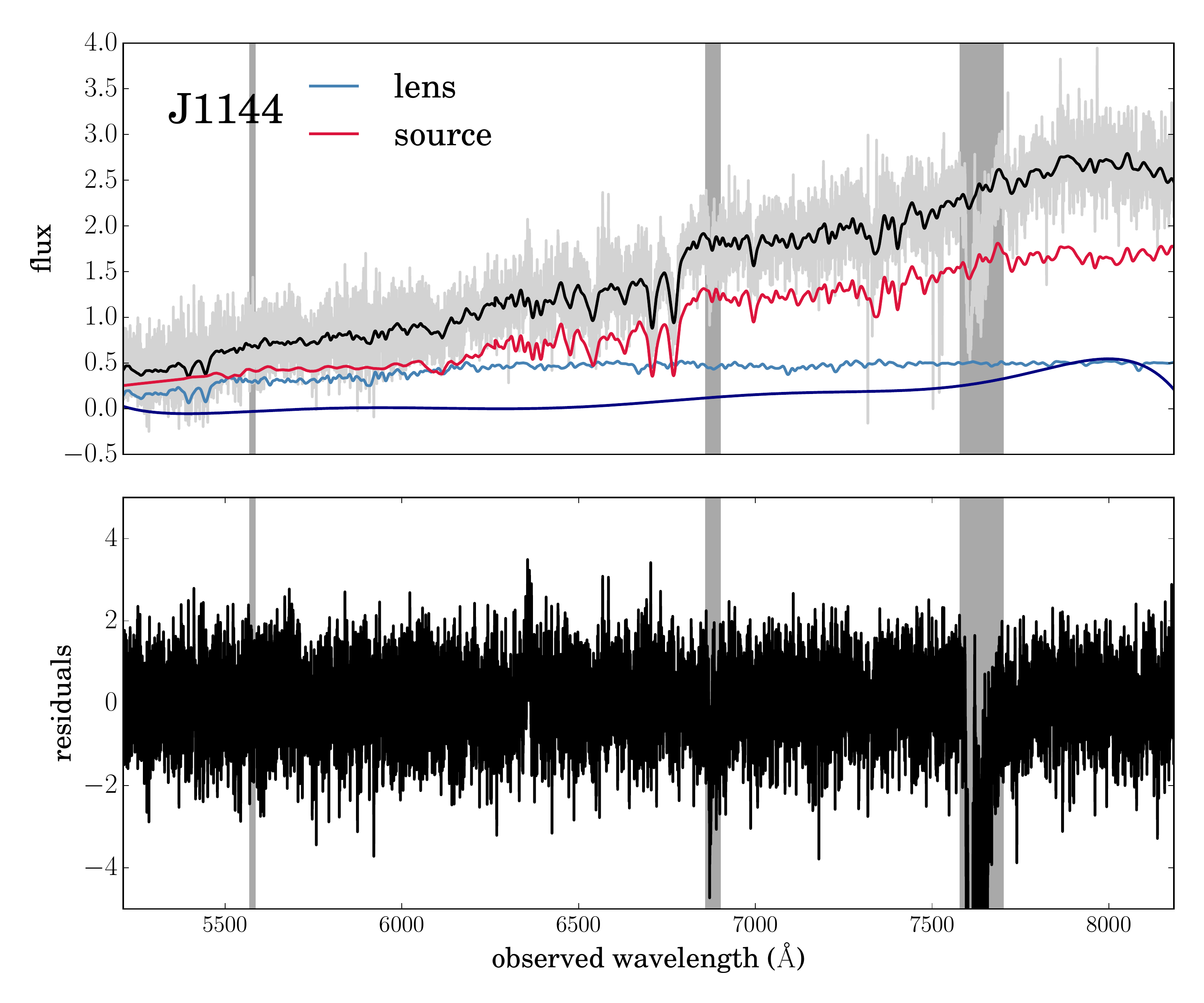}
\includegraphics[trim=20 20 20 20,clip,width=0.49\textwidth]{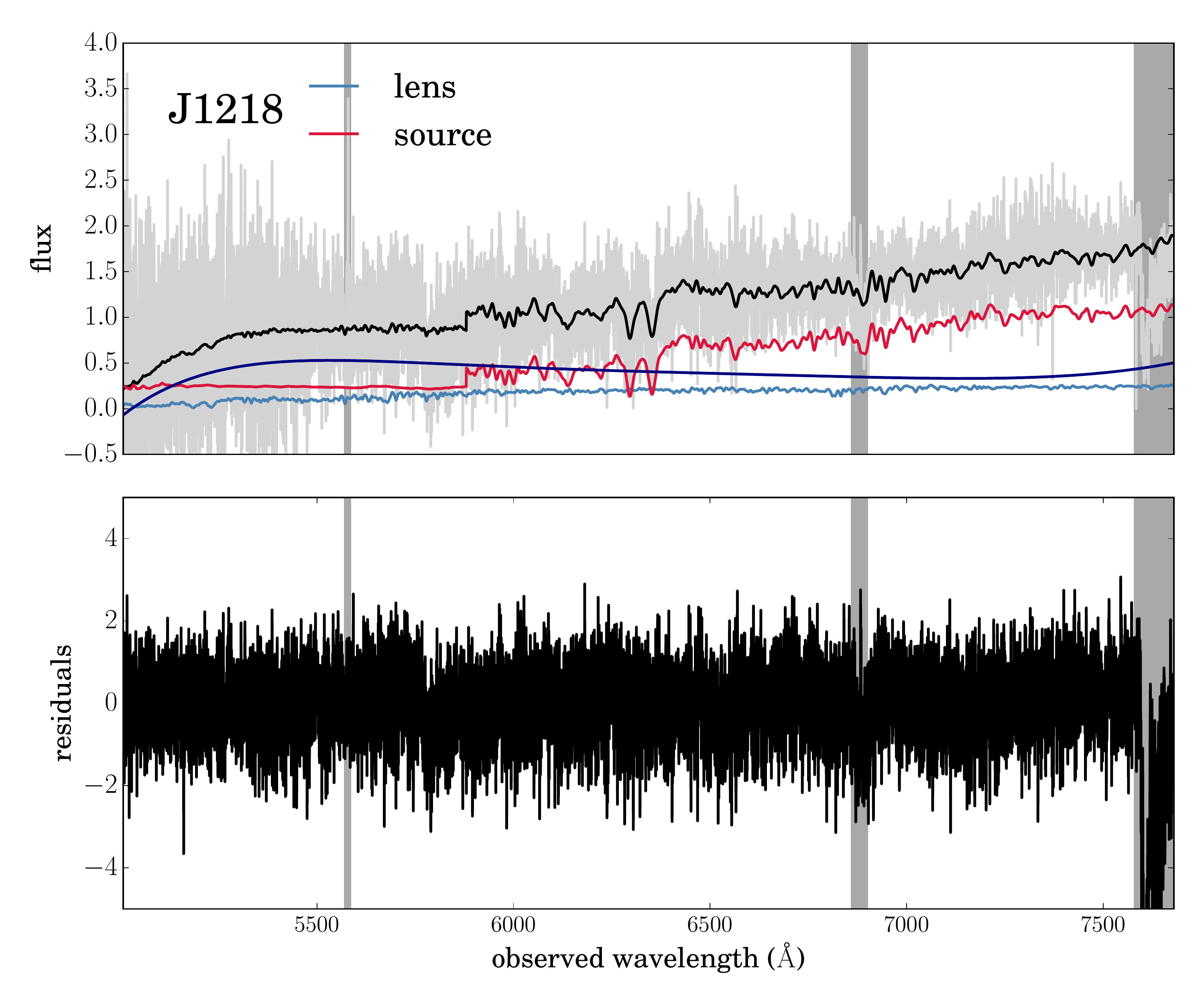}
\caption{Keck/ESI spectra for the 12 EELs in our sample, with kinematic models overplotted. In the upper panels, the data are shown in grey, the total model spectrum in black, and the contributions to the model from lens, source and continuum in blue, red and purple respectively. The lower panels show the residuals of the kinematic models; vertical grey bands represent regions that were masked from the fit due to the presence of telluric absorption features.}
\end{figure*}

\begin{figure*}
\subfigure{\includegraphics[trim=20 20 20 20,clip,width=0.49\textwidth]{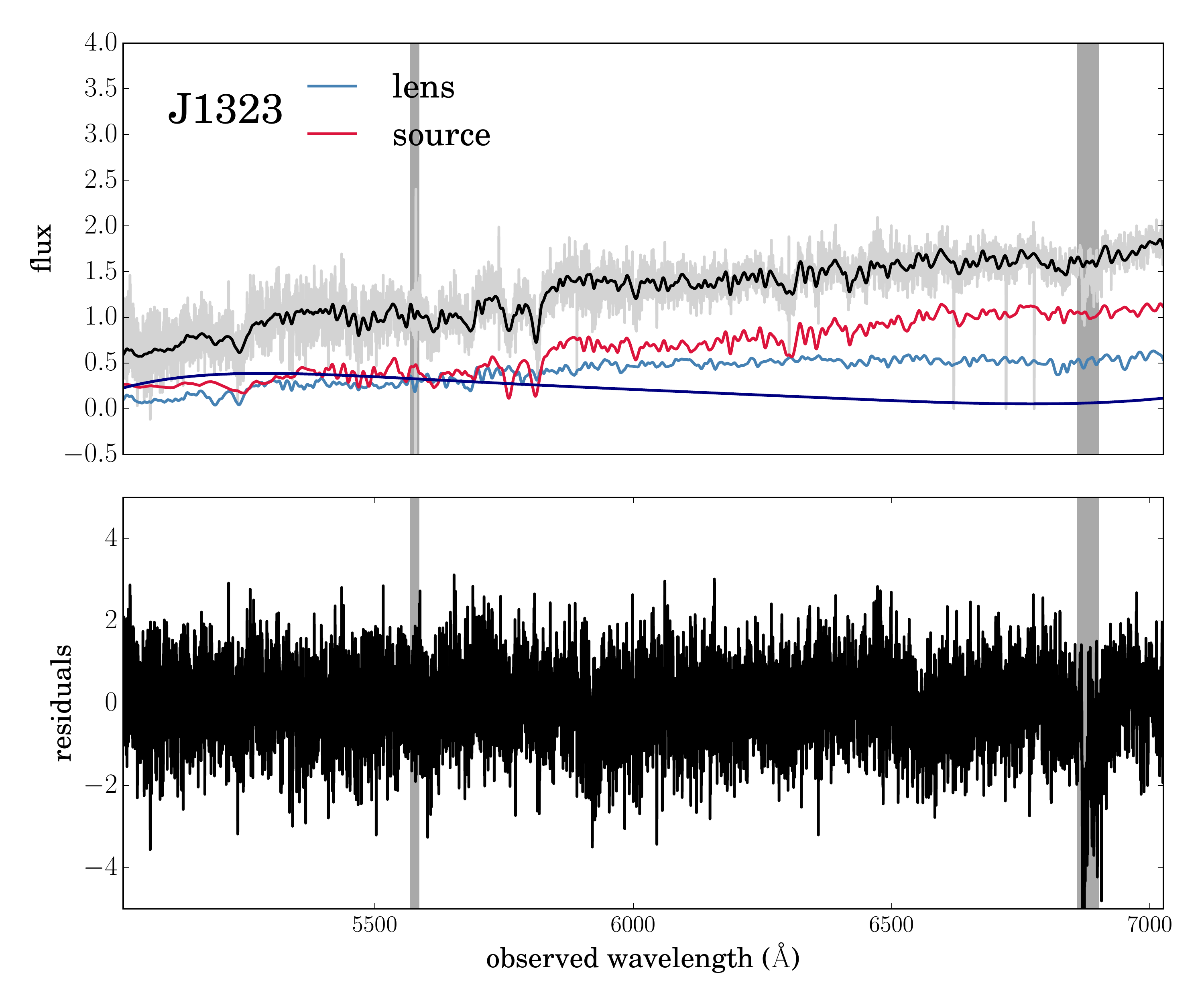}}\hfill
\subfigure{\includegraphics[trim=20 20 20 20,clip,width=0.49\textwidth]{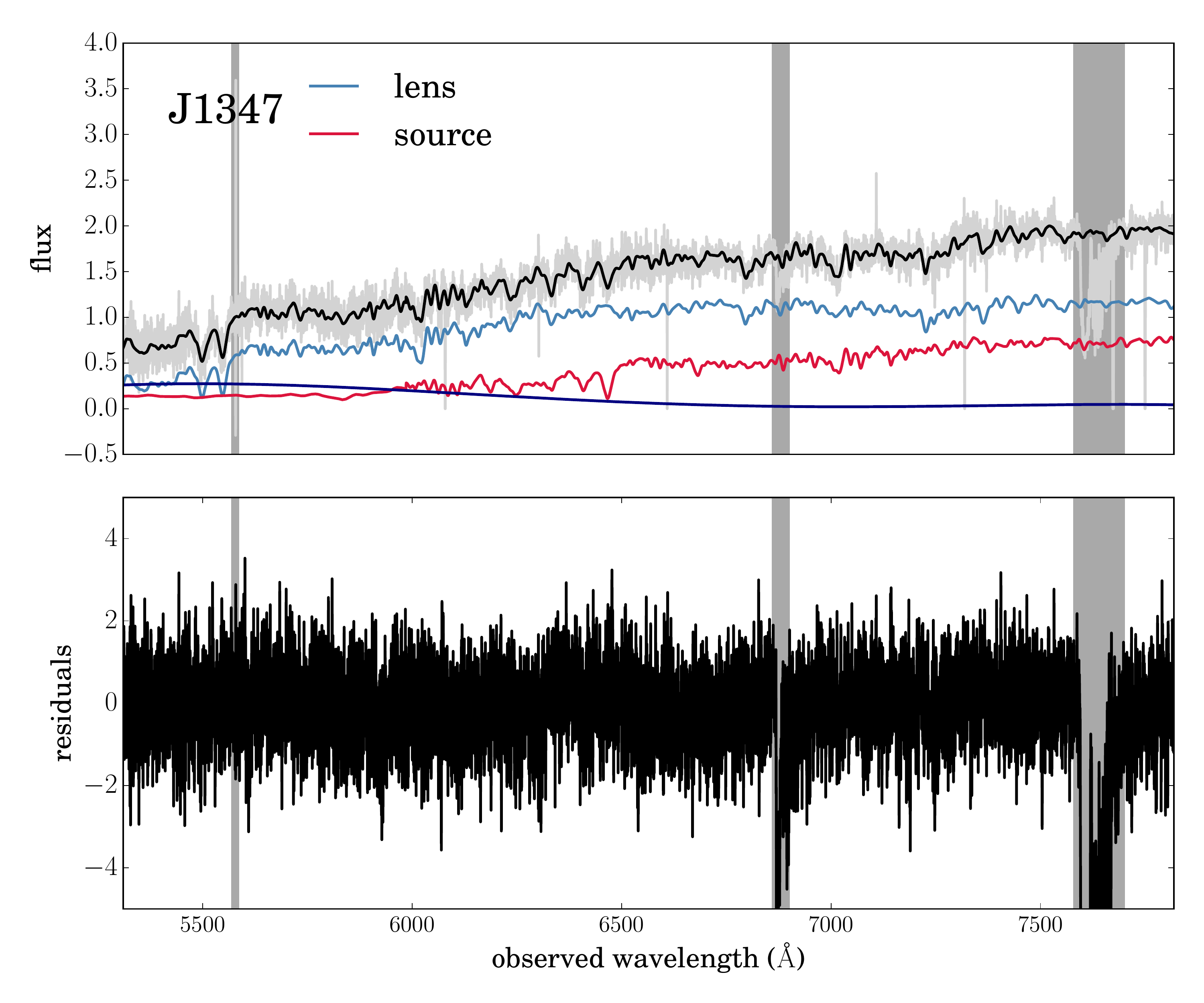}}\hfill

\subfigure{\includegraphics[trim=20 20 20 20,clip,width=0.49\textwidth]{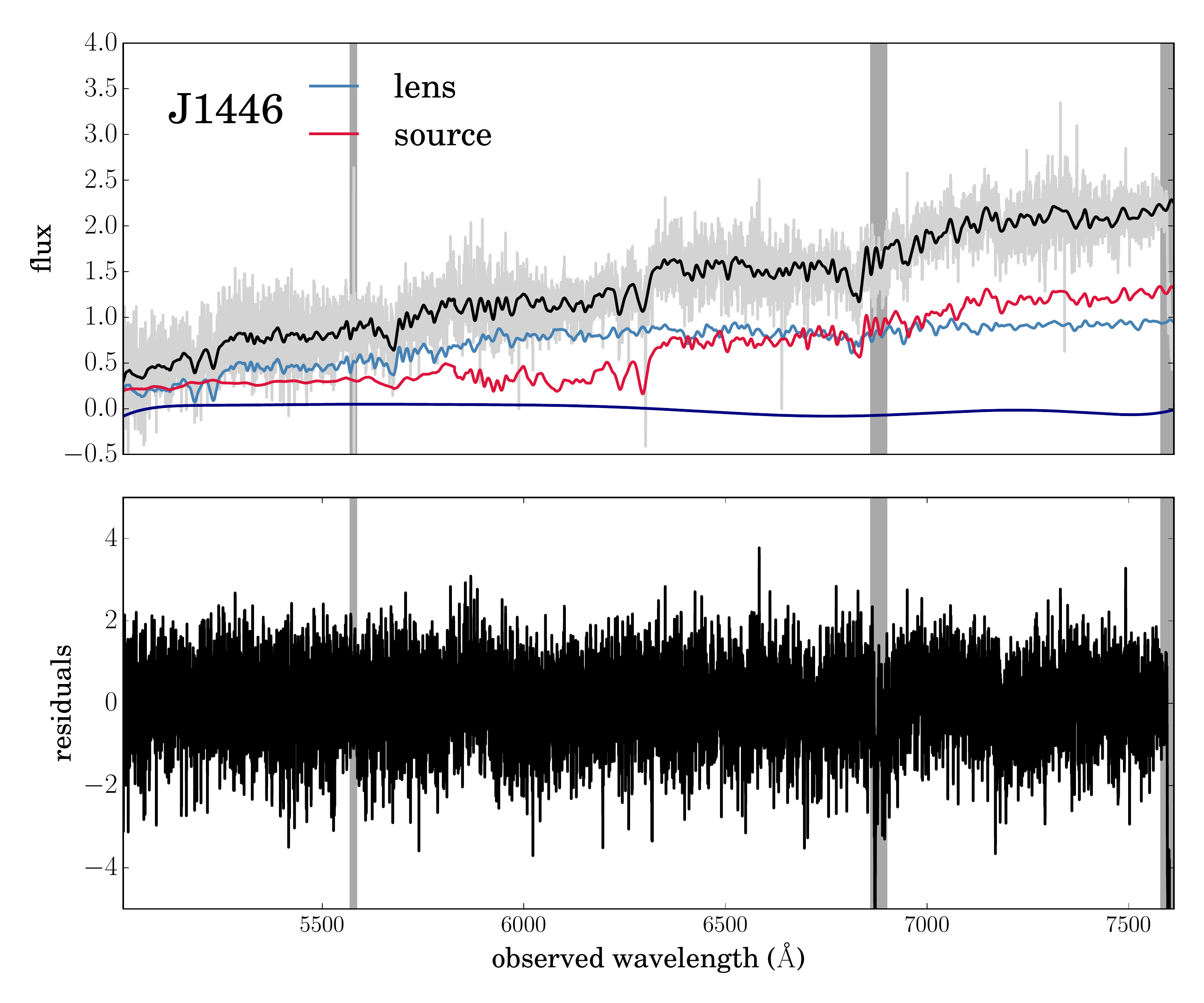}}\hfill
\subfigure{\includegraphics[trim=20 20 20 20,clip,width=0.49\textwidth]{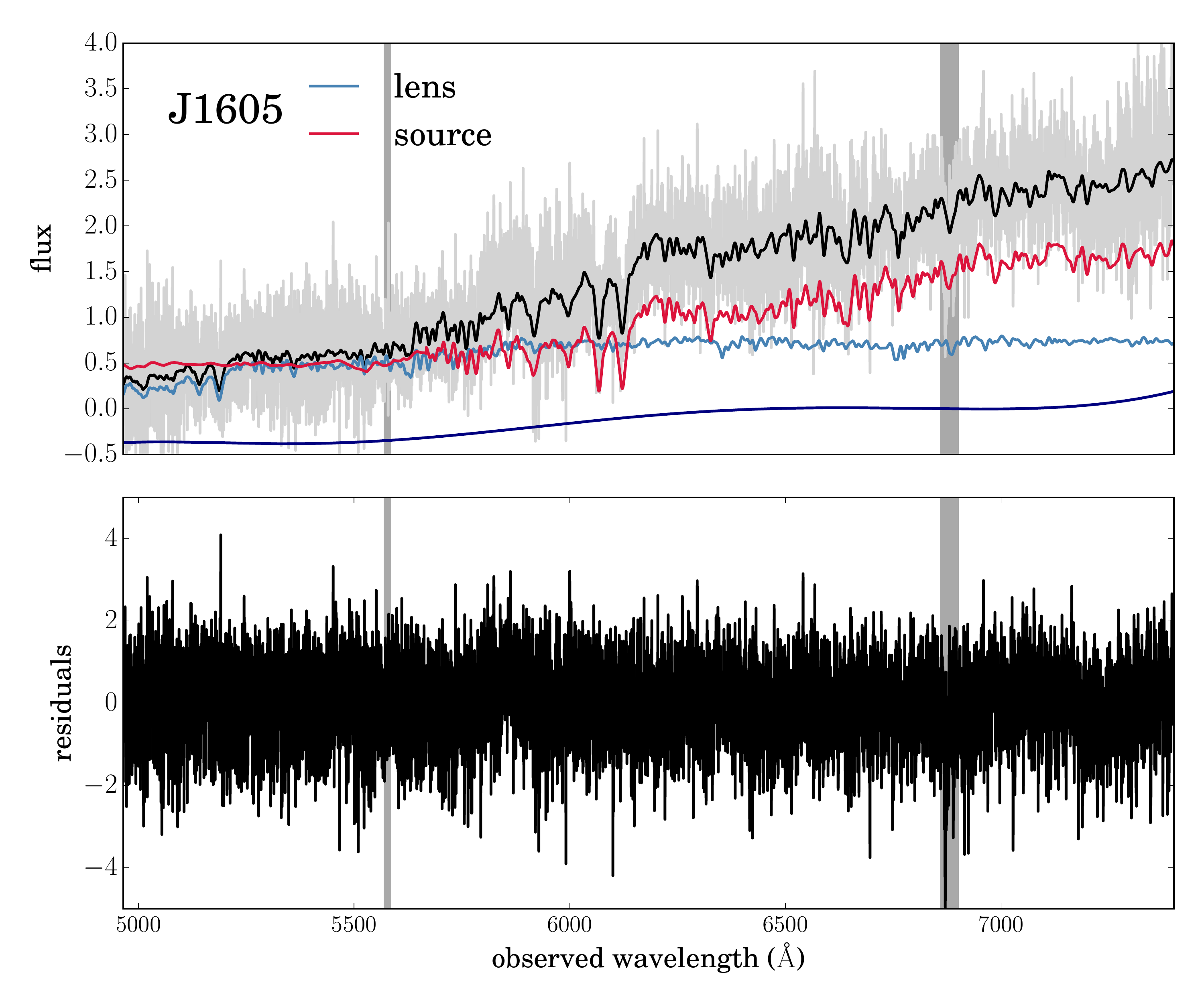}}\hfill

\subfigure{\includegraphics[trim=20 20 20 20,clip,width=0.49\textwidth]{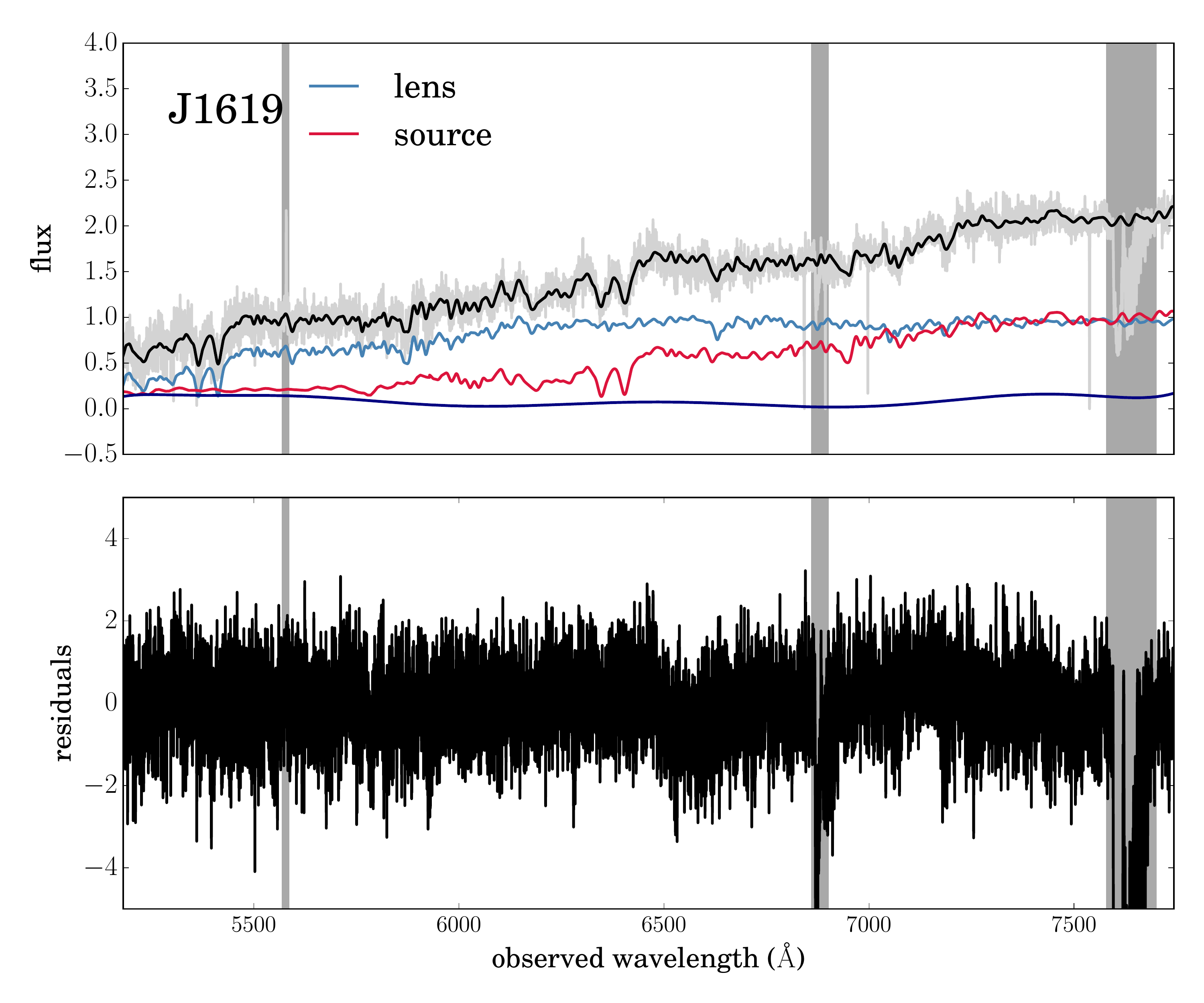}}\hfill
\subfigure{\includegraphics[trim=20 20 20 20,clip,width=0.49\textwidth]{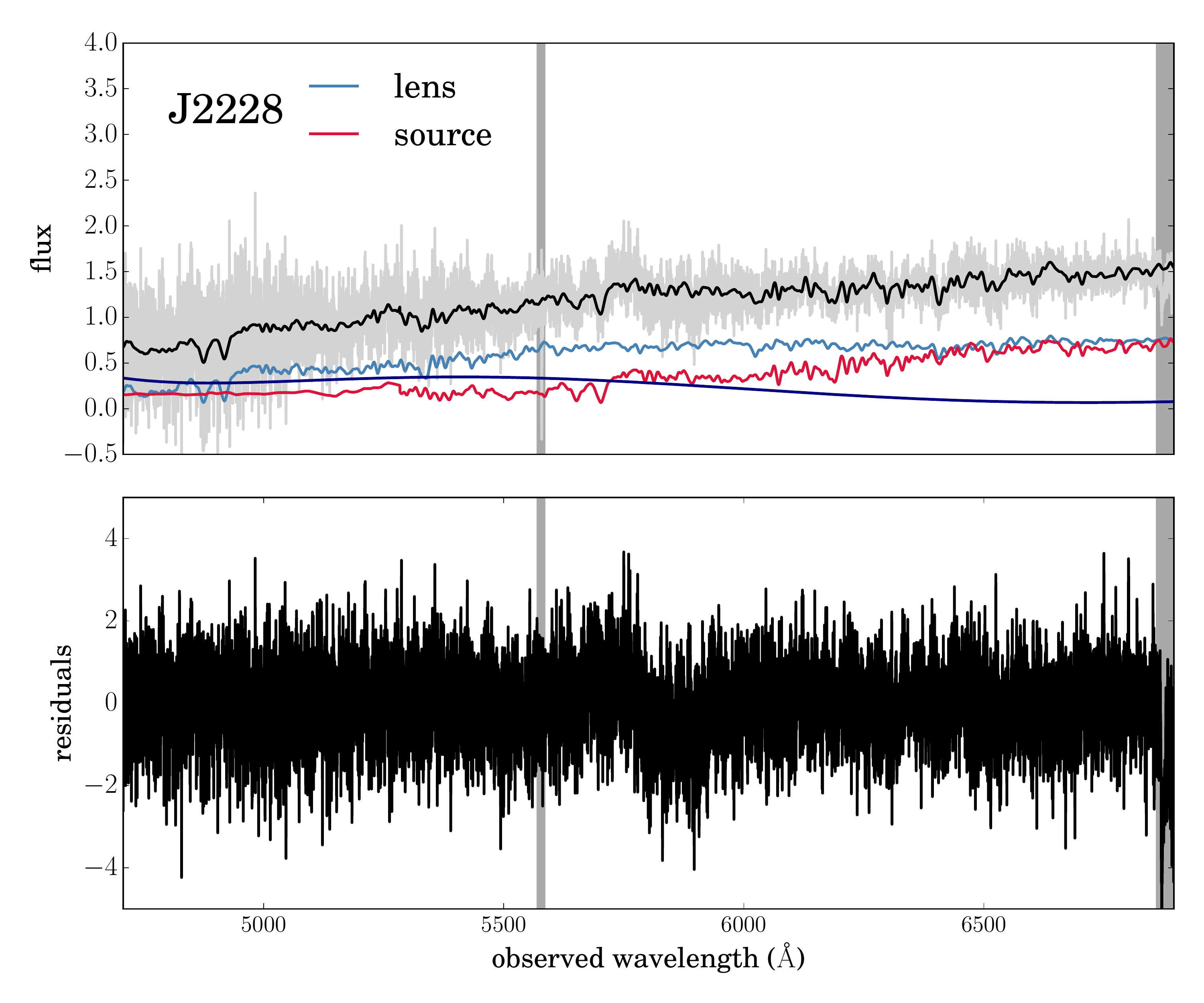}}\hfill

\contcaption{}
\end{figure*}

\begin{table}
 \centering
\begin{tabular}{cccC{1.9cm}}\hline
 EEL & $R_e$ (kpc) & $\sigma$ (kms$^{-1}$) & $\log I_e$ (L$_{\odot}$ kpc$^{-2}$) \\\hline
J0837+0801 & 4.46 $\pm$ 0.05 & 253.27 $\pm$ 13 & 9.14 $\pm$ 0.05 \\
J0901+2027 & 3.25 $\pm$ 0.03 & 205.40 $\pm$ 10 & 8.88 $\pm$ 0.04 \\
J0913+4237 & 4.05 $\pm$ 0.04 & 192.56 $\pm$ 10 & 8.86 $\pm$ 0.04 \\
J1125+3058 & 1.17 $\pm$ 0.00 & 182.27 $\pm$ \phantom{0}9 & 9.62 $\pm$ 0.01 \\
J1144+1540 & 9.61 $\pm$ 0.03 & 224.83 $\pm$ 11 & 8.53 $\pm$ 0.02 \\
J1218+5648 & 6.79 $\pm$ 0.04 & 191.08 $\pm$ 10 & 8.68 $\pm$ 0.03 \\
J1323+3946 & 2.36 $\pm$ 0.02 & 162.17 $\pm$ \phantom{0}8 & 9.14 $\pm$ 0.05 \\
J1347$-$0101 & 5.38 $\pm$ 0.07 & 151.95 $\pm$ \phantom{0}8 & 8.35 $\pm$ 0.02 \\
J1446+3856 & 1.59 $\pm$ 0.01 & 206.60 $\pm$ 10 & 9.49 $\pm$ 0.01 \\
J1605+3811 & 2.56 $\pm$ 0.01 & 160.40 $\pm$ \phantom{0}8 & 9.08 $\pm$ 0.02 \\
J1619+2024 & 5.23 $\pm$ 0.03 & 283.10 $\pm$ 28 & 8.95 $\pm$ 0.03 \\ 
J2228$-$0018 & 4.15 $\pm$ 0.03 & 149.18 $\pm$ 15 & 8.81 $\pm$ 0.03 \\\hline
\end{tabular}
\caption{FP data for the EELs source galaxies. Photometry is evaluated in the rest-frame Johnson $V$ band using the photometric models presented in \protect\citet{Oldham2016}, and kinematics are measured as described in Section 2.}
\end{table}

\section{Fundamental plane}

\subsection{The observed fundamental plane}

\begin{figure*}
 \centering
  \includegraphics[trim=20 20 20 20,clip,width=0.49\textwidth]{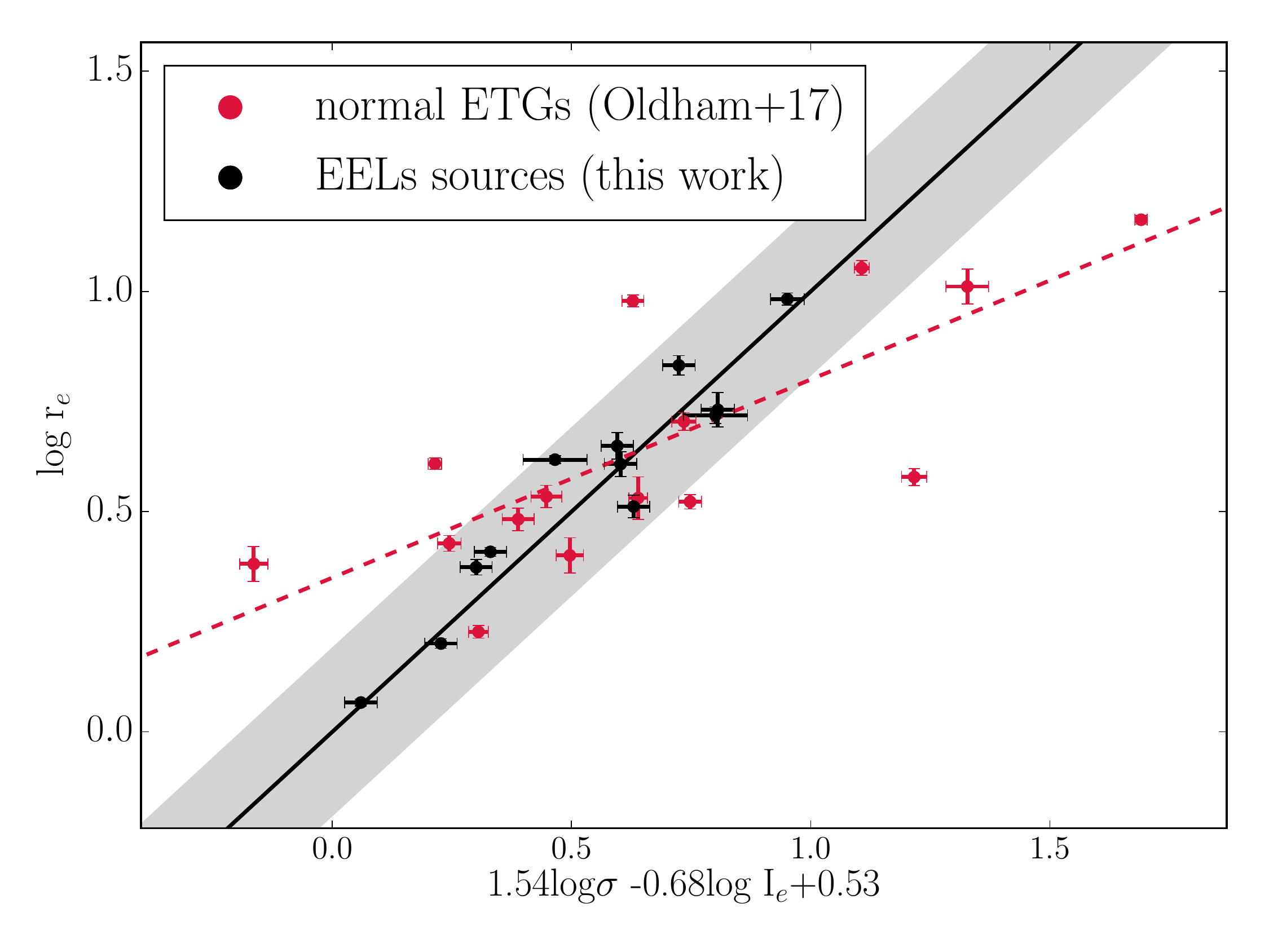}
\includegraphics[trim=20 20 20 20,clip,width=0.49\textwidth]{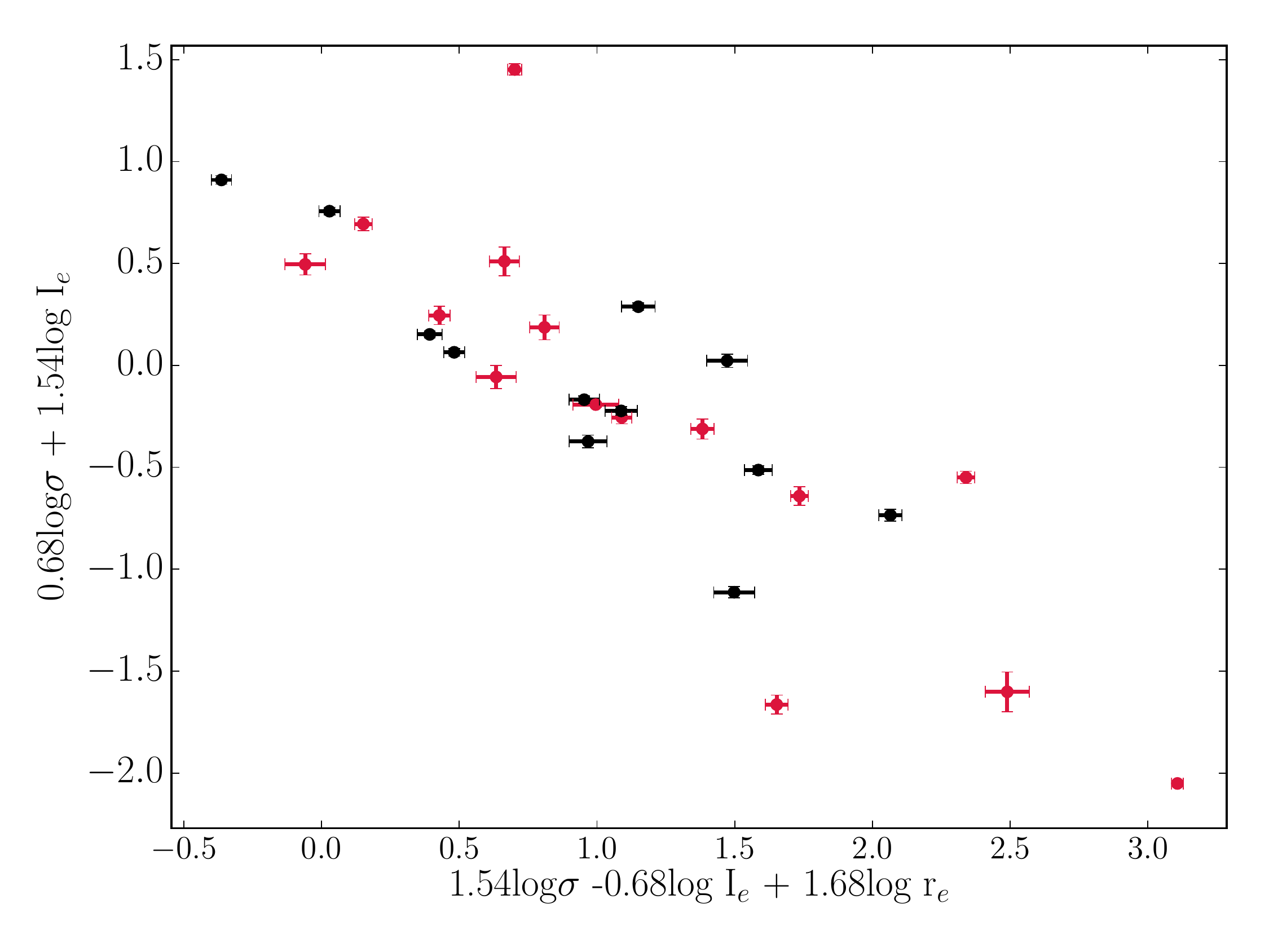}
\caption{The observed FP of the massive compact EELs sources, shown in side-on (left) and face-on (right) projections, with the similar-redshift, less compact ETGs from \citet{OldhamHD} shown in red for comparison. The FP is steeper in $\log \sigma$ and shallower in $\log I_e$ relative to that of the `normal' \citet{OldhamHD} galaxies, which indicates that the both the dark and stellar mass structure and stellar populations vary as a function of the plane parameters. The FP also has a comparable intrinsic scatter to that of normal ETGs, implying that the intrinsic scatter is driven by stellar population properties. Velocity dispersions, surface brightnesses and effective radii are measured in units 200kms$^{-1}$, $10^{9}$L$_{\odot}$ and kpc respectively.}
\end{figure*}

\begin{table*}
 \centering
\begin{tabular}{cC{1.3cm}C{1.6cm}C{1.3cm}C{1.6cm}C{1.6cm}C{1.3cm}C{1.3cm}C{1.3cm}C{1.6cm}}\hline
 & $\alpha$ & $\beta$ & $\gamma$ & $\mu_{\sigma}$ & $\mu_{I_e}$ & $\tau_{\sigma}$ & $\tau_{I_e}$ & $\rho$ & $\log\sigma_{FP}$ \\\hline
\multirow{2}{*}{Observed} & $1.54 \pm 0.36$& $-0.68 \pm 0.06$ & $0.53 \pm 0.02$ & $\phantom{0}0.00 \pm 0.03$ & $-0.03 \pm 0.14$ & $0.07 \pm 0.02$ & $0.38 \pm 0.11$ & $0.19 \pm 0.32$ & $-1.31 \pm 0.19$\\
 & $1.76 \pm 0.29$ & $-0.69 \pm 0.04$ & $0.56 \pm 0.01$ & $-0.02 \pm 0.02$ & $-0.04 \pm 0.13$ & $0.06 \pm 0.02$ & $0.39 \pm 0.11$ & $0.07 \pm 0.31$ & -- \\\hline
\multirow{2}{*}{Intrinsic} & $1.41 \pm 0.35$ & $-0.66 \pm 0.06$ & $0.56 \pm 0.02$ & $-0.02 \pm 0.03$ & $-0.02 \pm 0.12$ & $0.08 \pm 0.02$ & $0.35 \pm 0.10$ & $0.03 \pm 0.30$ & $-1.28 \pm 0.18$ \\
 & $1.69 \pm 0.28$ & $-0.69 \pm 0.04$ & $0.56 \pm 0.01$ & $-0.01 \pm 0.02$ & $-0.03 \pm 0.12$ & $0.07 \pm 0.02$ & $0.37 \pm 0.11$ & $0.07 \pm 0.30$ & -- \\\hline
\end{tabular}
\caption{Parameters and associated uncertainties for the fundamental plane, modelled according to the equation $\log R_e = \alpha \log\sigma + \beta \log I_e + \gamma$, with intrinsic scatter $\sigma_{FP}$ about $\log R_e$. The independent variables $\log \sigma$ and $\log I_e$ are modelled as being drawn from Gaussian distributions with mean $\vec{\mu} = (\mu_{\sigma}, \mu_{I_e})$ and variance $\vec{\tau}^2 = ((\tau_{\sigma}^2, \rho \tau_{\sigma}\tau_{I_e}), (\rho \tau_{\sigma}\tau_{I_e}, \tau_{I_e}))$. Velocity dispersions, surface brightnesses and effective radii are measured in units 200kms$^{-1}$, $10^{9}$L$_{\odot}$ and kpc respectively. The first and second rows represent models with and without an intrinsic scatter; allowing for an intrinsic scatter makes the FP shallower in the $\log\sigma$ directions; the third and fourth rows represent the same, but after accounting for the bias introduced by the EELs source selection function. The selection has a negligible effect on the FP parameters.}
\end{table*}


We model the FP for the twelve EELs sources as
\begin{equation}
 \log R_e = \alpha \log \sigma + \beta \log I_e + \gamma
\end{equation}
for effective radius $R_e$ in kpc, velocity dispersion $\sigma$ in units of 200 kms$^{-1}$ and effective surface brightness $I_e$ in $10^9$ L$_{\odot} / \textrm{kpc}^2$.

We follow the formalism presented by \citet{Kelly2007} by modelling both $\log \sigma$ and $\log I_e$ as being drawn from multivariate Gaussian distributions with some mean $\mu$ and dispersion $\vec{\tau}$, and allowing some intrinsic scatter $\sigma_{FP}$ along the direction of $\log R_e$; regarding the ongoing discussion on how to fit the FP \citep{Hyde2009}, we note that this method is closer to the case of minimising the residuals along $\log R_e$ rather than perpendicular to the plane. Full details on the form of the likelihood function can be found in \citet{Kelly2007}, but essentially, we construct the likelihood for the data given a particular set of plane and parent distribution parameters, and obtain the posterior distribution using an MCMC exploration of the parameter space. The FP that we infer is shown in both side-on and face-on projections in Figure 2, and our inference on the parameters is summarised in Table 2. As our sample is small, we test the robustness of our inference by remodelling random subsamples of the dataset and find that the uncertainty introduced by this is small, and less than half the size of the statistical uncertainties. We find that the observed FP of the EELs sources is steep relative to the FP of `normal' galaxies, with $\alpha = 1.54 \pm 0.36$ as compared to the $\alpha = 1.24 \pm 0.07$ that is found locally, and $\beta = -0.68 \pm 0.06$ compared to the local value $\beta = -0.82 \pm 0.02$ \citep{Jorgensen1995}; note though that both are consistent with the `normal' FP parameters within $2\sigma$. As a reference we use the sample of 17 ETGs belonging to the $z=0.545$ cluster MACSJ0717.5+3745 which were presented in \citet{OldhamHD}; these ETGs have a similar redshift and mean velocity dispersion to those in the EELs sample, but FP parameters that are consistent with the local values. Figure 2 shows that these do indeed fall differently on the FP. We also note that \citet{Auger2010a} found that the inclusion of an intrinsic scatter around the FP further tilts the plane of normal ETGs along the $\log\sigma$ direction to give $\alpha = 1.02 \pm 0.20$ relative to the \citet{Jorgensen1995} value. We also find this to be the case, with the removal of an intrinsic scatter from our model giving $\alpha = 1.76 \pm 0.29$ (and having negligible effect on $\beta$). 

The fact that these ETGs have compact light profiles means that their luminosity-weighted velocity dispersions probe only their very central regions, where the dark matter fraction is minimal. To zeroth order, we might then expect that any deviation in their FP from the virial prediction must be due to variations in the stellar mass and dynamical structure across the plane, as opposed to variations in the dark matter. That we find their FP to be tilted relative to both the local ETG population \textit{and} the virial plane indicates that the properties of both the stellar mass \textit{and} the dark matter vary across the plane. However, before drawing conclusions from this, we must account for the effect of the selection function of the EELs sources on the FP parameters. 

\subsection{The intrinsic fundamental plane}

The EELs sources form a biased population, as they were found in a lens search, and this may have a non-trivial effect on the orientation of the FP. We now follow the reasoning presented in \citet{Oldham2016} to correct for this bias and so recover the \textit{intrinsic} FP of these compact systems.

The selection function of the EELs sources has three main contributions. Firstly, the source must be lensed by the foreground object; this relates to the cross-section for lensing. Secondly, the inclusion of an EEL in the SDSS spectroscopic sample requires the lens+source system as a whole to satisfy the criteria of the SDSS target selection \citep{Strauss2002}, which itself is non-trivial, though the main effect here is that the system is bright. Finally, the EEL must pass our spectroscopic search, which is subjective but imposes criteria such as the (magnified) source galaxy flux being comparable to the lens galaxy flux and the redshifts of the two objects approximately satisfying $0.1 \lesssim z \lesssim 0.7$. The combination of these conditions leads to some selection function that modifies the intrinsic population of compact galaxies to the population of EELs sources that we observe.

Of these three contributions, the latter two are difficult to quantify and should not introduce any large bias into our measurement of the FP, although they will push us to the high-surface-brightness end. On the other hand, the first -- the lensing cross section -- introduces a selection function such that we are relatively more efficient at selecting compact galaxies at lower velocity dispersions. We can understand this as follows: differential magnification introduces a bias towards smaller objects (closer to the line-of-sight of the lens), whereas, for a given size, there is no bias as a function of luminosity, and therefore velocity dispersion (assuming the latter is dominated by stellar mass). The result of this is that an object of fixed velocity dispersion becomes increasingly likely to be seen in the lensed population relative to the intrinsic population as it becomes more compact.

This bias enters the FP in two ways, making its orientation biased towards objects with (a) small $R_e$ and (b) large $\log I_e$. Following \citet{Oldham2016}, we modify our likelihood for the FP data for the $i^{th}$ EEL, given a model, by a function describing the magnification achieved by the lens as a function of source size $F_{i}(R_{e,i})$, which serves as a proxy for the probability of the source being identified in the lens search. The result is summarised in Table 2 for models with and without intrinsic scatter, and shows that the bias is negligible -- probably because of the low intrinsic scatter, which makes selection effects unimportant -- and the FP of these compact systems remains tilted relative to that of normal ETGs. As mentioned in Section 3.1, this seems to indicate that the structure of both the dark and luminous mass components varies across the FP, i.e. as a function of mass. A discussion of this result is presented in Section 5.1.

\section{Physical models}

We can also combine our kinematic measurements, which give a measure of the \textit{total} mass in the central regions, with our photometric measurements from \citet{Oldham2016}, which give a measure of the \textit{stellar} mass, to make inference on the central dark matter fraction. By further supplementing these measurements with abundance matching relations, we can develop toy models to reconstruct the dark halo on large scales and so investigate the dark structure. We emphasise that our data are only sensitive to the mass in the \textit{central} regions, and that strong assumptions about how the stellar mass (which we can measure) relates to the virial mass (which we cannot measure) must be made. Nevertheless, abundance matching relations have been shown to be robust \citep{Behroozi2010}, and can therefore provide useful insight here, allowing us to probe the halo structure out to relatively high redshifts.

Our physical models are constructed as follows. For each EEL, we take the model of the lens system from \citet{Oldham2016} and use the surface brightness profile $I(R)$ for the source to construct its stellar mass profile, assuming a stellar mass-to-light ratio which is set by the total magnitude and total stellar mass (under the assumption of either a Chabrier or a Salpeter IMF). We use a generalised Navarro-Frenk-White (gNFW) profile to construct the dark matter halo as
\begin{equation}
 \rho(r) = \frac{\rho_0}{\Big(\frac{r}{r_s}\Big)^{\xi} \Big( 1 + \frac{r}{r_s}\Big)^{3-\xi}}
\end{equation}
where the normalisation $\rho_0(M_{\star})$ is drawn from a normal distribution based on the $z=0.6$ $M_{halo}(M_{\star})$ abundance matching relations of \citet{Behroozi2013}; the inner density slope $\xi$ is drawn from a normal distribution $\mathcal{N}(\mu_{\xi}, \tau_{\xi}^2)$ (such that $\mu_{\xi}$ characterises the mean inner density slope of the EELs source population and $\tau_{\xi}$ represents their scatter) and the scale radius $r_s$ is either calculated from the mass-concentration relation \citep[][though we note that this relation was constructed assuming NFW haloes, and it is not clear that it should still be valid when the NFW assumption is relaxed]{Maccio2008} or modelled as $\mathcal{N}(\mu_{r_s}, \tau_{r_s}^2)$ and inferred along with the $\xi$ parameters. Though the NFW profile is commonly used due to its success in describing the dark haloes of galaxies in dark-matter-only simulations \citep{NFW1997}, the use of the gNFW is motivated by the expectation that baryonic physics should modify the halo in some way -- either contracting the halo via adiabatic processes, creating a super-NFW cusp $\xi > 1$ \citep{Blumenthal1986,Gnedin2004}, or hollowing it out via heating from active galactic nuclei or dynamical friction, leading to a sub-NFW central region $\xi < 1$ \citep[e.g.][]{ElZant2004,Mashchenko2006,Governato2012,Laporte2012,Velliscig2014}. 

We then use the total density profile, $\rho(r) = \rho_{DM}(r) + \rho_{\star}(r)$, to calculate a velocity dispersion using the spherical Jeans equation
\begin{equation}
 \frac{d}{dr}(l\sigma_r^2) + 2\frac{\beta(r)}{r}l\sigma_r^2 = l(r) \frac{GM(r)}{r^2}
\end{equation}
where $l(r)$ is the luminosity density of the stars (i.e. the deprojected surface brightness profile, assuming axisymnmetry), $\beta(r) = 1 - \sigma_t^2 / \sigma_r^2$ is the anisotropy parameter and $\sigma_r(r)$ the radial velocity dispersion, which we then project along the line of sight as
\begin{equation}
 \frac{1}{2}I(R)\sigma_{los}(R)^2 = \int_R^{\infty} \frac{l \sigma_r^2 r \mathrm{d}r}{\sqrt{r^2 - R^2}} - R^2 \int_R^{\infty} \frac{\beta l \sigma_r^2 \mathrm{d}r}{r \sqrt{r^2 - R^2}}
\end{equation}
to obtain the line-of-sight velocity dispersion $\sigma_{los}(R)$ as a function of projected radius $R$. Finally, we integrate $\sigma_{los}$ over a circular aperture of radius $1.5 R_e$ (comparable to the effective aperture over which the lensed source spectra were extracted) to obtain the velocity dispersion that would be measured within the aperture
\begin{equation}
 \sigma_{ap}^2 = \frac{\int_0^{R_{ap}} I \sigma_{los}^2 R \mathrm{d}R}{\int_0^{R_{ap}} I R \mathrm{d}R}
\end{equation}
\citep[see e.g.][]{Mamon2005}. We investigate both isotropic and constant-anisotropy models across the range $-2 \leq \beta \leq 1$, and find that the difference is sufficiently small that our data cannot distinguish between them; we therefore adopt $\beta = 0$ in all models. Thus, given the luminous structure of an EELs source, we can reconstruct the halo and compare the implied central (dark+light) mass with the central velocity dispersion to constrain the dark matter structure in the inner regions. 

Our inference is shown in Figure 3 and Table 3. We find that our inference on $\xi$ does not depend significantly on whether we impose the mass-concentration relation or allow $\xi$ and $r_s$ to vary freely; this suggests that the mass-concentration relation does not break down for deviations of the halo profile from NFW of the scale that we are considering here. Moreover, we find that our inference on the halo inner slope depends strongly on the IMF that is assumed (though the scale radius does not), with bottom-heavy (Salpeter) and bottom-light (Chabrier) IMFs requiring haloes that are excavated and contracted respectively. This is a common problem in attempts to disentangle the dark and luminous mass structure of ETGs and in general must be broken through the use of multiple tracer populations or other mass probes \citep[e.g.][]{Newman2013,OldhamAuger2016b}. In this case, the construction of stellar population models from spectral absorption features would provide an independent probe of the IMF which would allow this degeneracy to be broken, and will be the topic of a future work. However, we also note that high-mass ETGs have been shown to require Salpeter-like IMFs in general \citep[e.g.][]{Auger2010b, vanDokkum2010, Cappellari2012}; though the EELs sources span a range in mass, we note that their masses are generally high, which here would imply the existence of sub-NFW haloes and the removal of dark matter from their central regions. We discuss this investigation further in Section 5.2.

\begin{figure}
 \centering
\includegraphics[trim=10 20 20 20,clip,width=0.48\textwidth]{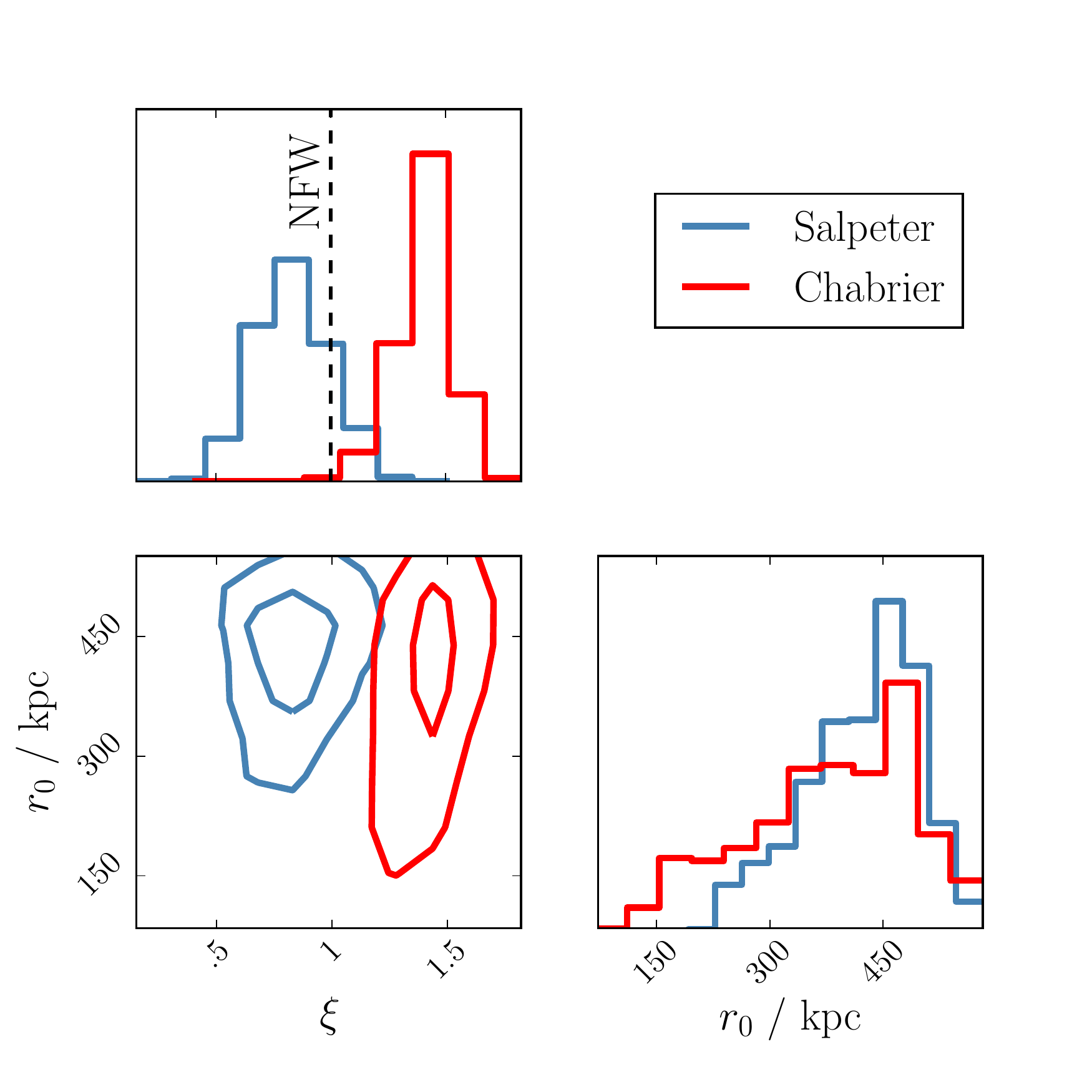}
\caption{Inference on the structure of the dark halo for Chabrier (red) and Salpeter (blue) IMFs, compared to the expected inner slope from an NFW profile ($\xi = 1$; black dashed line). Whilst inference on the scale radius is robust, the measurement of the inner slope depends strongly on the form for the IMF that is assumed, with a bottom-heavy (Salpeter) IMF requiring less dark matter centrally than the NFW prediction and a bottom-light (Chabrier) IMF requiring more dark matter than the NFW prediction. An independent IMF probe is needed if we are to robustly distinguish between these scenarios; however, assuming that the EELs sources have Salpeter-like IMFs, consistent with previous IMF studies, implies the presence of sub-NFW haloes and points towards the action mergers and accretion in removing dark matter from the central regions.}
\end{figure}

\begin{table}
 \centering
\begin{tabular}{cccccc}\hline
IMF & free & $\mu_{\xi}$ & $\tau_{\xi}$ & $\mu_{r_s}$ & $\tau_{r_s}$  \\\hline
\multirow{2}{*}{Chabrier} & $\xi$ & $1.23_{-0.09}^{+0.08}$ & $0.29_{-0.08}^{+0.05}$ & -- & -- \\
& $\xi$, $r_s$ & $1.32_{-0.10}^{+0.18}$ & $0.37_{-0.11}^{+0.12}$ & $323_{-38}^{+79}$ & $72_{-46}^{+10}$ \\\hline
\multirow{2}{*}{Salpeter} & $\xi$ & $0.71_{-0.13}^{+0.13}$ & $0.45_{-0.08}^{+0.12}$ & -- & -- \\
& $\xi$, $r_s$ & $0.91_{-0.15}^{+0.13}$ & $0.54_{-0.1}^{+0.09}$ & $443_{-23}^{+35}$ & $61_{-21}^{+10}$ \\\hline
\end{tabular}
\caption{Parameters and associated uncertainties for physical models of the dark matter mass structure, assuming a halo density profile as given in Equation 2, with (a) free $\xi$ and $r_s$ determined from the mass-concentration relation, and (b) free $\xi$ and $r_s$, and assuming (a) Chabrier and (b) Salpeter IMFs. A Chabrier IMF requires the presence of more dark matter centrally than the NFW+abundance matching prediction; conversely, a Salpeter IMF requires less dark matter than the NFW+abundance matching prediction.}
\end{table}

\section{Discussion}

\subsection{The fundamental plane}

In Section 3, we found the FP of the EELs sources to be only marginally consistent with that of local ETGs, being relatively steep and relatively shallow in the $\log \sigma$ and $\log I_e$ directions respectively. This tells us about both the stellar mass structure of massive compact galaxies, and the astrophysics behind the FP.

The steepness of the FP in the $\log \sigma$ direction follows from the fact that the EELs sources form an extremely massive and compact population: first, their central regions are very stellar-mass-dominated, and second, any luminosity-weighted property is only able to probe their innermost parts. Thus the velocity dispersion that we measure is dominated by the stellar mass, as opposed to the dark mass, and any rotation of the FP from the virial prediction must be driven by variations in their stellar mass structure and dynamics as a function of mass. The fact that their FP appears to be rotated towards, but still inconsistent with, the virial plane, implies that the non-homology in the dark matter structure of ETGs cannot be the only cause of the well-established FP tilt, and that mass-dependent variations in the stellar component -- for instance, in the kinematics, mass structure or IMF -- must also be important. 

On the other hand, the shallowness of the FP in the $\log I_e$ direction implies that the relation between the effective radius and effective surface brightness of massive compact ETGs differs from that of normal ETGs. In \citet{Oldham2016}, we showed that the EELs sources have bugle+envelope-like structures, consistent with a two-phase picture of their growth in which the formation of a compact core at early times is followed by the accretion of material at larger radii at later times, eventually causing the compact cores to grow into the larger ETGs that we see at low redshifts. In this picture, the EELs sources are at a relatively early stage in their evolution such that the accreted envelope remains pronounced, unlike more evolved ETGs, which appear more spheroidal. This suggests a natural interpretation for the tilt in the FP that we observe along $\log I_e$ as the result of a difference in the luminous mass structure of massive compact ETGs.

We also note the difference in the FP of these compact ETGs from that of cluster ETGs at similar redshifts. In \citet{OldhamHD}, we presented the FP of massive ETGs in the cluster MACSJ0717 at $z=0.545$; this sample has the same mean velocity dispersion as the EELs sources ($\mu_{\sigma} \sim 200$ kms$^{-1}$), but their FP is extremely consistent with that of local ETGs, with no evidence for a tilt (see Figure 2). The main implication of this result is that these cluster ETGs have already experienced the majority of their structural evolution by $z\sim0.5$; contrary to the EELs sources, which have remained compact. In \citet{Oldham2016}, we also showed that the EELs sources do not appear to occupy unusually dense environments. Taken together, this pair of results is interesting as it provides further evidence for the accelerated growth of galaxies in dense environments: if any massive compact galaxies existed in MACSJ0717, they would be some of the brightest ETGs in the cluster and so should have been included in that study. The fact that they are not suggests that all the initially-compact ETGs in that dense environment have already evolved into larger systems like those we see locally, whereas the EELs sources, in less dense environments, remain at an earlier stage of their evolution. To make a more quantitative statement here would require a thorough characterisation of the selection functions of these two samples; nevertheless, this direct comparison strongly highlights the differences between these two populations.

Finally, we measure the FP to have a small scatter, $\log \sigma = -1.31 \pm 0.19$, comparable to that of the normal FP (\citealt{Auger2010a} find the SLACS lenses to have $\sigma = 0.049 \pm 0.009$). If scatter in the dark and the light structure contributed equally to the overall FP scatter, then we would expect to find a reduced intrinsic scatter in the FP of the EELs sources, where the low dark matter fractions should make the contribution to the scatter from the dark matter minimal. The fact that the intrinsic scatter is not decreased when variations in the dark matter are effectively removed in this way implies that this scatter is largely due to variations in the stellar structure and stellar populations, and that the dark matter structure in ETGs -- at least in their central regions -- is subject to less variation. Equally, though, our sample size is small, which makes statistical properties of the sample, such as the intrinsic scatter, subject to uncertainty.

\subsection{Physical models}

In Section 4, we constructed toy models for the mass structure in the EELs sources in order to probe their dark matter distributions. Since we cannot currently constrain the IMFs in these systems, we constructed these models assuming a universal IMF which is either Chabrier and Salpeter, and found that the inference on the halo structure is (perhaps unsurprisingly) very sensitive to the choice of IMF. If a Chabrier-like IMF is assumed, we find the sources to require \textit{more} dark matter in the centre than implied by the abundance matching+NFW prediction, whereas the use of a more bottom-heavy Salpeter IMF requires \textit{less} dark matter centrally than the abundance matching+NFW prediction. These two results are very different and imply correspondingly different physics: whilst a sub-NFW dark halo would imply the action of either strong AGN feedback or merger processes in removing dark matter from the centre, a super-NFW halo would indicate an important role for initial adiabatic processes during galaxy assembly. It might be possible to distinguish between these scenarios by using higher-signal-to-noise spectra to independently constrain the IMF in these systems via stellar population modelling, though the difficulty here would be in robustly measuring the continuum, given that the spectra contain the light from both the lens and the source galaxies. This is something we will explore in a future work. 

For now, however, we combine a number of recent developments in our understanding of ETGs to sketch a feasible picture of the structure and evolution of the galaxies in our sample. At the outset, we emphasise our assumption that all the galaxies in our sample have the same IMF: in reality, their mass range may imply a variation in the IMF across the sample, which would complicate the following picture; nevertheless, given that the non-universality of the IMF remains not well understood and that we cannot constrain the IMF from our data, we adopt this as a reasonable first model.

First, existing evidence suggests that ETGs of comparable stellar mass to the EELs sources require bottom-heavy IMFs \citep[e.g.][]{Auger2010b, vanDokkum2010, Cappellari2012}, and further, that the IMF of ETGs may be most bottom-heavy in their central regions \citep{MartinNavarro2015,OldhamInPrep}. Seen in the context of the inside-out growth scenario, in which ETGs grow in two stages by the formation of a compact core at high redshifts, followed by the accretion of lower-mass systems at large radii to form an extended envelope, this suggests that the formation of the initial core may occur in fundamentally different star formation conditions from the lower-mass systems that it subsequently accretes. In this picture, the EELs sources, which seem to have experienced very little accretion so far, should have very bottom-heavy IMFs, comparable with the most central regions of local ETGs. As shown in Figure 3, this would then imply that their halo profiles are sub-NFW. What is interesting about this is that one of the most promising mechanisms for the removal of dark matter from the inner regions of ETG haloes is dynamical friction from infalling satellites during merger events, which has been shown in simulations to be effective at transferring energy from the infalling objects to the halo and therefore causing the latter to expand \citep{Laporte2012}. 

In \citet{Oldham2016}, we revealed evidence for ongoing merger activity in the \emph{luminous} structure of these systems in the form of faint envelopes surrounding their compact cores; the possibility that they also have sub-NFW haloes may therefore present further evidence for their growth by mergers and accretion. This independent evidence, based on their \emph{dark} structure, provides further insight into their evolution and presents new potential evidence for the importance of dry merging in the growth of ETGs.

\section{Conclusions}

We have presented and modelled spectra and the resulting stellar kinematics for 13 early-type/early-type lenses (EELs) to explore the nature of the fundamental plane (FP) and the dark and light mass structure of the source galaxies. Our main conclusions are as follows.

\begin{enumerate}
 \item Relative to normal ETGs, the FP of the EELs sources is tilted towards, but still inconsistent with, the virial plane. Since the EELs sources are compact systems with small effective radii, their luminosity-weighted kinematics probe the very central regions where the dark matter fraction is low; this means their FP is sensitive predominantly to their \textit{luminous} material, as opposed to their \textit{dark matter}. The fact that their FP is rotated relative to both the FP of normal galaxies and the virial plane indicates that the properties of both the stellar populations and the dark matter are responsible for the well-known FP tilt, i.e. vary as a function of galaxy mass.
 \item The intrinsic scatter of the FP of the EELs sources is small ($\log \sigma = -1.31 \pm 0.19$), but comparable to that of the FP of normal galaxies. This implies variations in the inner dark matter structure do not contribute significantly to the scatter, which must instead be driven by scatter in the stellar properties.
 \item The halo structure can only be constrained, in the context of well-motivated models, when a universal IMF is assumed. However, on the basis of mounting evidence that massive, compact ETGs should have bottom-heavy IMFs, these galaxies appear to be best characterised by a dark matter halo with a sub-NFW inner density slope. This is consistent with a picture in which these galaxies are growing by successive mergers and accretion of smaller objects, and may therefore provide further evidence for their inside-out growth.
\end{enumerate}

\section{Acknowledgements}

LJO thanks Peter Behroozi for providing the $M_{halo}(M_{\star})$ tables that were used in the construction of the mass models. LJO thanks the Science and Technology Facilities Council (STFC) for the award of a studentship. MWA also acknowledges support from the STFC in the form of an Ernest Rutherford Fellowship. CDF acknowledges support from STScI (HST-GO-13661) and from the NSF (AST-1312329). LVEK is supported in part through an NWO-VICI career grant (project number 639.043.308). 

This paper includes data obtained at the W.M. Keck Observatory, which is operated as a scientific partnership among the California Institute of Technology, the University of California and the National Aeronautics and Space Administration. The Observatory was made possible by the generous financial support of the W.M. Keck Foundation. The authors wish to recognize and acknowledge the very significant cultural role and reverence that the summit of Mauna Kea has always had within the indigenous Hawaiian community.  We are most fortunate to have the opportunity to conduct observations from this mountain.

\FloatBarrier

\end{document}